\documentclass[11pt]{aastex}
\usepackage[twocolumn]{emulateapj5}
\usepackage{psfig,graphicx}
\def\gcrt {GCRT\, J1745-3009}
\def\dblpsr {J0737-3039}

\shorttitle{GCRT J1745-3009: a double neutron star binary ?}

\shortauthors{Turolla, Possenti, \& Treves}

\begin{document}

\title{Is the Bursting Radio-source GCRT J1745-3009 a Double Neutron Star Binary ?}

\author{R. Turolla\altaffilmark{1}, A. Possenti\altaffilmark{2}, A.
Treves\altaffilmark{3}}

\altaffiltext{1}{Department of Physics, University of
Padova, via Marzolo 8, 35131 Padova, Italy; turolla@pd.infn.it}
\altaffiltext{2}{Istituto Nazionale di Astrofisica, Osservatorio 
Astronomico di Cagliari, loc. Poggio dei Pini, Strada 54, 09012 Capoterra, 
Italy; possenti@ca.astro.it}
\altaffiltext{3}{Department of Physics and Mathematics, Universit\'a
dell'Insubria, Via Valleggio 11, 22100, Como, Italy;
treves@mib.infn.it}


\begin{abstract}
\gcrt \ is a peculiar transient radio-source in the direction of the Galactic Center. 
It was observed to emit a series of $\sim 1$~Jy bursts at 0.33 GHz, with typical 
duration $\sim 10$~min and at apparently regular intervals of $\sim 77$~min. If the
source is indeed at the distance of the Galactic Center as it seems likely, we show
that its observational properties are compatible with those
expected from a double neutron star binary, similar to the double pulsar system \dblpsr.
In the picture we propose the (coherent) radio emission comes from the shock originating in
the interaction of the wind of the more energetic pulsar with the magnetosphere of
the companion. The observed modulation of the radio signal is the consequence
of an eccentric orbit, along which the separation between the two stars varies. This 
cyclically drives the shock inside the light cylinder radius of the less energetic pulsar.
\end{abstract}



\section{Introduction}

Recently \cite{hym05} announced the remarkable discovery
of a powerful bursting radio-source located about $1\deg$ south of
the Galactic Center (GC). \gcrt \ was detected in 2002 during a radio
transient monitoring of the GC region at 0.33 GHz, and appears
to emit a series of bursts of typical duration $\sim 10$~min. The flux in a
single burst is $\sim 1$~Jy and no emission is detected
between the events to an upper limit of $75$~mJy. The
bursts seem to follow a regular pattern, with a time separation
of $\sim 77$~min. The bursting source is unresolved and no counterparts are found 
thus far in other spectral bands.

The quite high radio flux, combined with the limited spatial extent of
the emitting region derived from the decay time of the bursts,
implies a brightness temperature of $T_{b}\sim 10^{16}$~K placing
the source at the GC distance. If this is the case,
a coherent mechanism is required to power the observed radio
emission (\citealt{hym05}).

A number of possible interpretations for the peculiar properties of  \gcrt \
have been examined by \cite{hym05}. Their main conclusion is that, although
the option that this is a close-by radio-source (a very cold dwarf, or
an extrasolar planet) can not be dismissed, it appears much more likely that
\gcrt \ is indeed located close to the GC. Along this line, they consider
both a radio-pulsar and a neutron star-neutron star binary as possible options.
The case for a pulsar nature of \gcrt \ is more thoroughly discussed in a recent paper
by \cite{zx05}.

If one interprets the interburst time as due to orbital
motion, according to the original suggestion by
\cite{hym05}, the system may be similar to the double pulsar
\dblpsr \ (\citealt{burgay03}; \citealt{lyne04}).
In this {\em Letter} we further explore the possibility that \gcrt \ is
a double neutron star system, following the proposal by \cite{tt04}, that the
continuous emission from \dblpsr \ could be due to coherent
radiation, produced at the shock formed by the interaction of the relativistic
wind of the more energetic pulsar with the magnetosphere of the companion.

\section{The model}

We develop a simplified model for \gcrt \ by assuming that: i) the
source is at the distance of the GC, $D=8.5$~kpc, ii) the interval between the
bursts is an orbital period, $P=77$~min, and iii) the binary system consists of
two neutron stars (NS). Eight examples of NS+NS binaries in
which one of the star is a radio pulsar are presently known. Among these
\dblpsr \ is unique since both its components are detected as pulsars.
Moreover \dblpsr \ is the only NS+NS binary
($P_{orb}=2.4$~h, $a\sim 9\times 10^{10}$~cm, $e\sim 0.09$; \citealt{lyne04}),
which exhibits a continuous radio emission ($F_\mathrm{1.4\, GHz}\sim 5$~mJy)
about two times stronger than the emission from the two pulsars (\citealt{lyne04}). 
The fact that unpulsed radio emission occurs in a double neutron star
binary is of particular relevance in connection with \gcrt \ and is at the basis
of the scenario we are proposing. So far it has not been possible to establish if
the continuous radio flux from \dblpsr \ exhibits an orbital modulation. Were the
continuous radio emission steady or weakly modulated, despite the alleged analogies,
the two systems must differ in some respects, as discussed further on.

The origin of the continuous emission from \dblpsr \ is still
uncertain. The starting point is assumed to be the interaction of
the wind of the most luminous pulsar A ($P_{A}=0.02$~s,
$\dot E_{A}\sim 6\times 10^{33} \ {\rm erg\, s}^{-1}$,
$B_{A}\sim 6\times 10^{9}$~G) with the magnetosphere of pulsar B
($P_{B}=2.77$~s, $\dot E_{B}\sim 2\times 10^{30} \
{\rm erg\, s}^{-1}$, $B_{B}\sim 10^{12}$~G~\footnote{Actually, the magnetic field
of pulsar B could be somewhat smaller than this value, see \cite{lyut04}}). A
shock is produced when the energy density of the relativistic wind
of A equals the magnetic energy density of B. Since $\dot E_{A}\gg
\dot E_{B}$, the wind penetrates deep into the magnetosphere of B
and the shock is well within B's light cylinder radius at $r_{S}\sim
6\times 10^9$~cm  from B (\citealt{lyne04}; TT04). Under the reasonable
assumption that the shock is the site of particle acceleration,
TT04 suggested two possible scenarios for the continuous emission.
If the relativistic electrons originating in the shock fill a large
volume (size $\approx 10^{13}$--$10^{14}$~cm corresponding to $\approx 1\arcsec$
at the estimated distance of \dblpsr, $\sim 0.6$~kpc), optically thin synchrotron radiation
is able to account for the observed flux. On the contrary, in case the
size of the emitting region is comparable to that of the shock itself,
the only way to circumvent severe self-absorption is to consider coherent emission.

\begin{figure*}
\resizebox{17.cm}{!}{\includegraphics[angle=0,clip=]{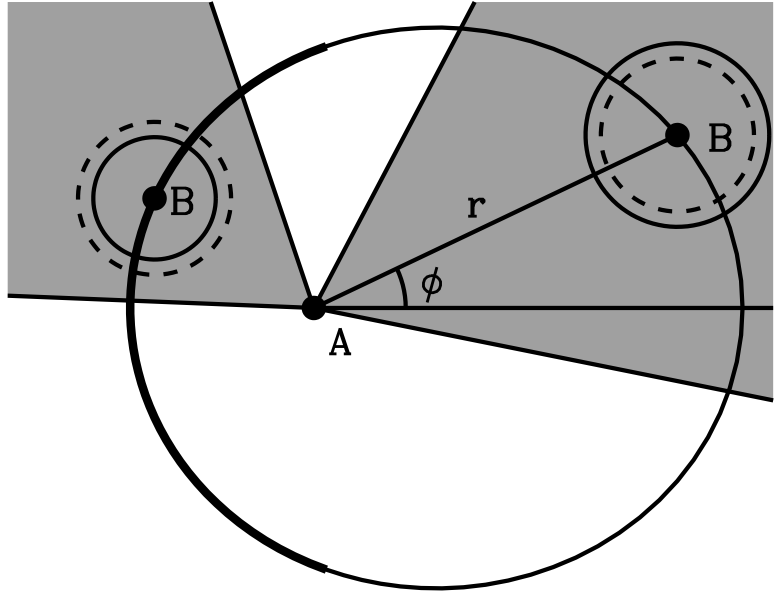}}
\vskip -2.cm
\caption{A cartoon of the putative double NS system in \gcrt. The orbit
of star B relative to star A has $e=0.4$. The dashed and full circles
represent the position of B's light cylinder and of the shock, respectively (not
to scale). Along the portion of the trajectory close to periastron and marked with
a heavy line the shock is inside the light cylinder. The shaded areas show the wind
of A (for illustrative purpose only).
\label{orbit}}
\end{figure*}

In the picture we propose for \gcrt \ the coherent emission
is activated when the relativistic wind of one component (which we call A)
penetrates the magnetosphere of the other (B), in analogy with what likely
occurs in \dblpsr. However, as already noticed and at variance with the case
of \dblpsr, the strong radio emission has a finite duration, $T\sim 10$~min. A possible
explanation is that the two NSs in \gcrt \ are in rather eccentric orbits, so that
the shock is inside B's light cylinder only when the separation
between the two stars drops below a limiting value. Clearly such a situation
will occur only close to periastron, as illustrated in Figure \ref{orbit}.

\begin{figure}
\resizebox{13.cm}{!}{\includegraphics[angle=0,clip=]{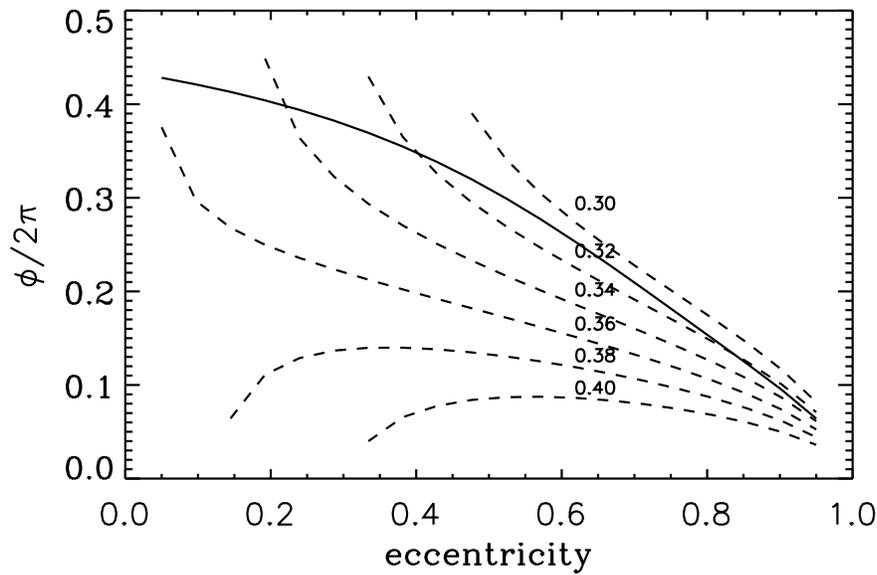}}
\caption{The solutions of eqs. (\ref{phi}) (full line) and (\ref{shock}) (dashed lines) as
a function of the eccentricity. Each dashed curve is labelled by the corresponding value
of the spin period $P_{B}$ in seconds.
\label{intersec}}
\end{figure}

The orbit semi-major axis follows from Kepler's law 

\begin{eqnarray}
\label{kepler}
a &=& 3.5\times 10^{10}
\left(\frac{M_{{tot}}}{M_\odot}\right)^{1/3}\left(\frac{P}{{\rm 1 h}}
\right)^{2/3}\, {\rm cm}\cr
& \sim &5.7\times 10^{10}\, {\rm cm}
\end{eqnarray}
where $M_{{tot}}= M_{A}+ M_{B}\sim 2.6M_\odot$ is the
total mass. The area swept by the radius in a
time $T$ is simply $S=\pi a^2\sqrt{1-e^2}(T/P)$, where $e$ is the
orbit eccentricity. From this it follows that the portion of the
orbit centered at the periastron and travelled in a time $T$ is
limited by the values of the phase angle $\phi$ and $2\pi-\phi$,
where $\phi$ as a function of $e$ is implicitly given by

\begin{equation}
\label{phi}
\int_\phi^\pi\frac{1}{(1-e\cos\phi)^2}\,d\phi=\frac{\pi}{(1-e^2)^{3/2}}
\frac{T}{P}\, .
\end{equation}
The solutions of eq. (\ref{phi}) for different values of $e$ and $T/P=10\,
\rm{min}/77\, \rm{min}=0.13$ are shown
in Figure \ref{intersec} (full line).

The distance of the shock from B, $r_{S}$, is given by

\begin{equation}
\label{endens}
\frac{B_{B}^2}{8\pi}=\frac{\dot E_{A}}{4\pi\chi_{A}(r-r_{S})^2c}
\end{equation}
where $B_{B}=B_{B}(R_{B})(R_{B}/r_{S})^3$ assuming a dipole field, $R_{B}=10^6\, {\rm cm}$ is
the radius of star B, $B_{B}(R_B)$ is the polar surface field,
$\dot E_{A}$ is the wind luminosity
emitted by A in a solid angle $4\pi\chi_{A}$ and $r=a(1-e^2)/(1-e\cos\phi)$
is the orbital separation.

The burst radio luminosity at $\nu=0.33$~GHz is

\begin{equation}
\label{lr}
L_{R}\sim 4\pi\chi_{R} F_\nu\nu D^2\sim 2.7\times 10^{31}\chi_{R} \
{\rm erg\,s}^{-1}
\end{equation}
where  $\chi_{R}$ is the fraction of the total solid angle covered by the
coherent radio beam. We take $L_{R}$ to be proportional to the fraction of
$\dot E_{A}$ intercepted by the shock

\begin{equation}
\label{edota}
\chi_{R}L_{R}(\phi)=\frac{\pi\xi\dot E_{A}r_{S}^2}{4\pi
\chi_{A}(r-r_{S})^2}\, ;
\end{equation}
$\xi$ represents the efficiency for the production of (coherent) radio photons.
Note that the radio
luminosity which appears in the previous equation depends on the orbital phase through $r$ and
$r_{S}$.
To derive an estimate of $\dot E_{A}$, we evaluate eqs. (\ref{endens}) and (\ref{edota})
at the periastron, where $\phi=\pi$, $r=a(1-e)\sim 3\times 10^{10}\, {\rm cm}$, and
we take $L_{R}(\pi)=L_{R}$, the latter as given by eq. (\ref{lr}). This
results in

\begin{eqnarray}
\label{edotaval}
\dot E_{A} & \sim & 1.8\times 10^{35}\left[\frac{B_{B}(R_{B})}{10^{12}\,
{\rm G}}\right]^{-1}\left(\frac{\chi_{A}}{0.1}\right)
\left(\frac{\chi_{R}}{0.1}\right)^{3/2}\cr
&&\times \left(\frac{\xi}{10^{-3}}\right)^{-3/2}\ {\rm erg\,s}^{-1}\, .
\end{eqnarray}
Inserting the previous expression for $\dot E_{A}$ into eq. (\ref{endens})
and imposing that the shock is at the light cylinder radius of B,
$r_{S}=r_{c, B}=cP_{B}/2\pi\sim 4.8\times 10^9 P_{B}\
{\rm cm}$, results in a further relation between the phase $\phi$ and $e$

\begin{eqnarray}
\label{shock}
\frac{1-e^2}{1-e\cos\phi}&=&8.4\times 10^{-2}P_{B}+20P_{B}^3
\left[\frac{B_{B}(R_{B})}{10^{12}\,
{\rm G}}\right]^{-3/2}\cr
&&\times\left(\frac{\chi_{R}}{0.1}\right)^{3/4}\left(\frac{\xi}{10^{-3}}\right)^{-3/4}.
\end{eqnarray}
The request that $r_{S}=r_{c,B}$ at a given point P guarantees that
$r_{S}<r_{c,B}$ at all points along the orbit which are closer than P to
periastron.

\begin{figure*}
\resizebox{8.1cm}{!}{\includegraphics[angle=0,clip=]{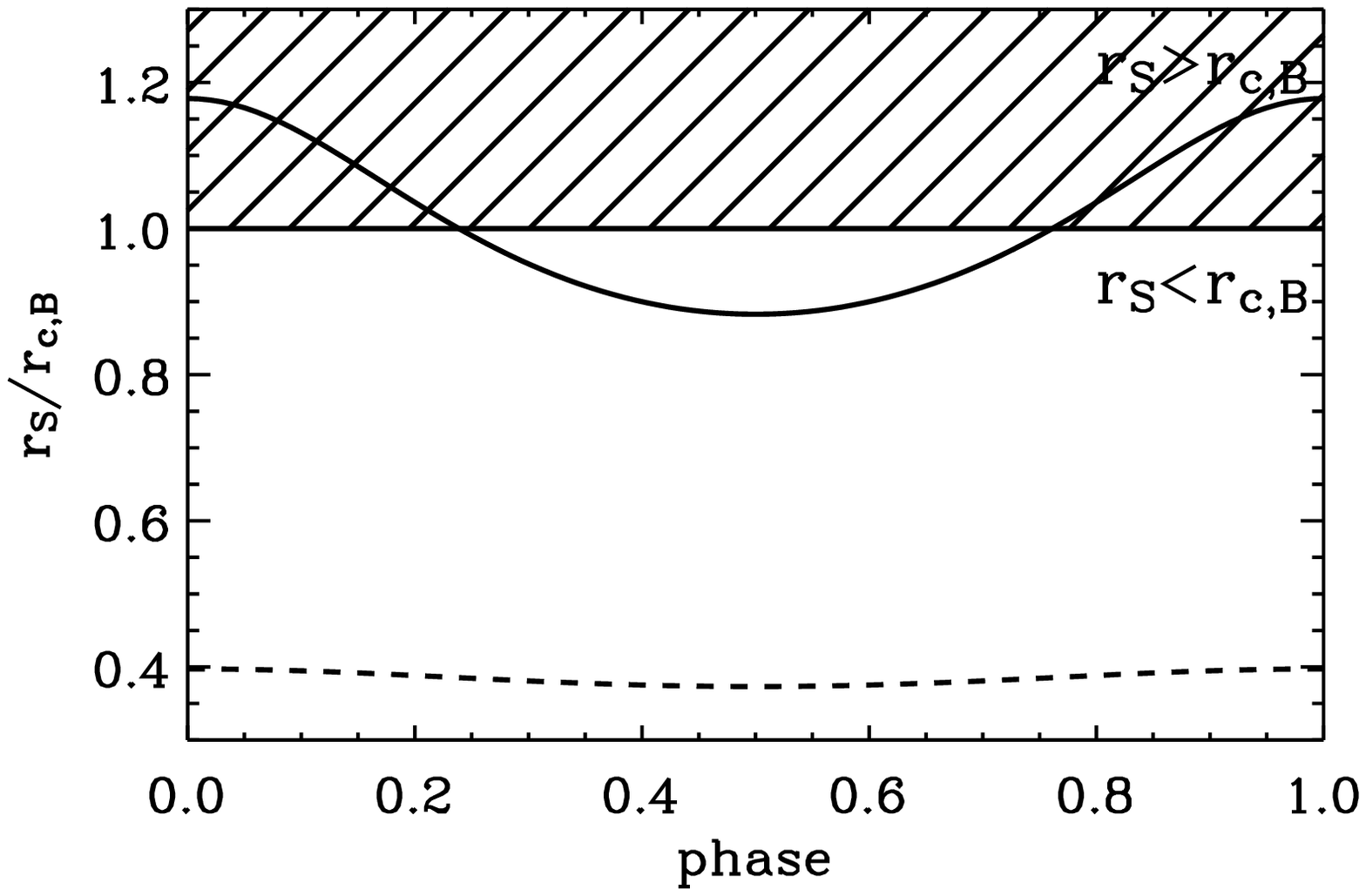}}\hspace{1mm}
\resizebox{8.1cm}{!}{\includegraphics[angle=0,clip=]{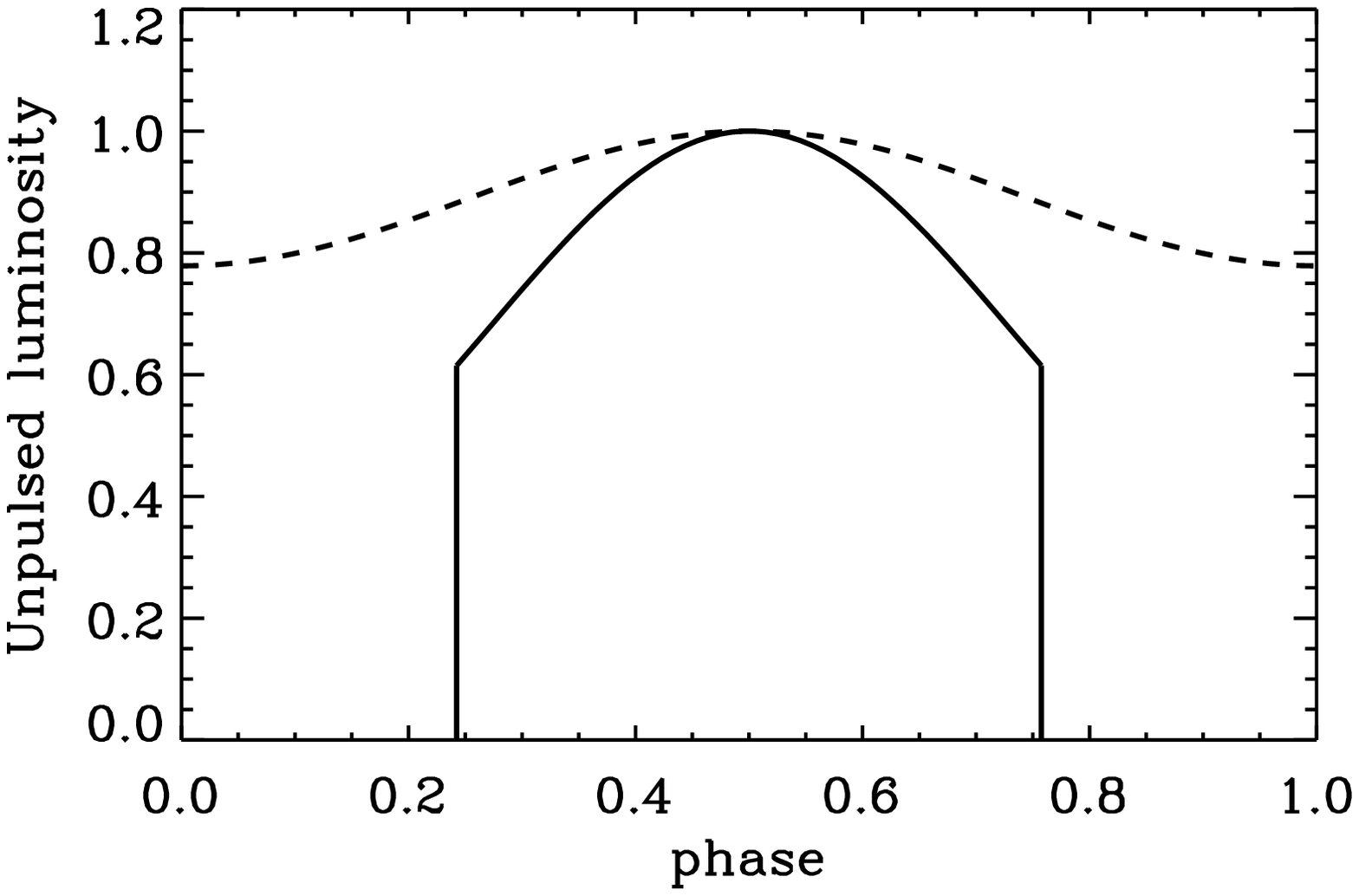}}
\caption{Left: The variation of the shock radius with the orbital phase. Actually,
a shock can form only outside the hatched region, where $r_{S}< r_{c,B}$.
Right: Same for the unpulsed radio luminosity (in arbitrary units; see text for 
details). The dashed lines show the same quantities for \dblpsr.\label{rs-lc}}
\end{figure*}

In order for our model to work, conditions (\ref{phi}) and (\ref{shock}) must be
satisfied together. Only in this case the shock is inside the light cylinder radius of B for
the observed duration of the burst $T$. The values of $\phi$ solutions of eq. (\ref{shock})
are shown in Fig. \ref{intersec} as a function of the eccentricity and
for different values of $P_{B}$ and $B_{B}(R_B)=10^{12}\, {\rm G}$;
all other parameters have been taken equal to their
reference values. As it can be seen from Fig. \ref{intersec}, the simultaneous fulfillment of
both conditions is possible only in a quite restricted range of spin periods of star B,
$0.31\ {\rm s} \la P_{B}\la 0.35 \ {\rm s}$. The corresponding values of the eccentricity
are in the range $\sim 0.3$--0.6. In an even tighter interval of periods, around
$P_{B}=0.32\ {\rm s}$, a second solution for $\phi$ is present at higher values of the eccentricity.
The allowed periods increase with $B_{B}$; however, for  $5\times 10^{11}\, {\rm G}< B_{B}(R_B)
< 3\times 10^{12}\, {\rm G}$ they are still in the 0.3--0.4 s range. Increasing (decreasing) the beaming
factor $\chi_{R}$ (the efficiency $\xi$) results in shorter periods.
The shock distance from star B, as derived from eq. (\ref{endens}),
is shown in Figure \ref{rs-lc} (left panel) for $e=0.4$. By combining eqs. (\ref{endens}) and (\ref{edota})
it easy to see that $L_{R}(\phi)\propto r_{S}^4$. The corresponding light curve
is shown in Figure \ref{rs-lc} (right panel); no radio emission occurs
when star B phase angle is outside the range for which $r_{S}\leq r_{c,B}$.
In Figure \ref{rs-lc} the orbital variation of $r_{S}$ and $L_{R}$ is shown
for the parameters typical of \dblpsr \ for comparison. In this case it is always
$r_{S}< r_{c,B}$. The shock distance changes only slightly along
the orbit because of the lower eccentricity and the orbital modulation of the unpulsed 
flux is $\sim 10\%$.

The wind luminosity of pulsar B can be estimated as

\begin{equation}
\label{edotb}
\dot E_{B} \sim 3\times 10^{32}\left[\frac{B_{B}(R_{B})}{10^{12}\,
{\rm G}}\right]^2\left(\frac{P_{B}}{0.35 \, \rm s}\right)^{-4} \ 
{\rm erg\, s}^{-1}\, .
\end{equation}
Although the ratio of the wind
luminosities $\dot E_{A}/\dot E_{B}$ is somewhat smaller than that inferred
in \dblpsr, it is still $\dot E_{B}\ll\dot E_{A}$ and
the energy density of the wind of A exceeds that of B by a factor $\sim 10$
at  B light cylinder.
This ensures consistency with our starting assumption about the shock
forming as the result of the interaction of the wind of A with the magnetosphere of B.

\section{Discussion}

The scenario outlined in the previous section shows that if one chooses for the
model parameters 
values close to those introduced above, the main observational
features of \gcrt \ can be successfully reproduced. Although the general picture we
propose resembles that of \dblpsr, the two binary systems differ in more than
one respect. In particular, the shock radiative efficiency needs to be higher in 
\gcrt, $\sim 10^{-3}$ as compared to $\sim 10^{-4}$ in \dblpsr \ (TT04). Lower values 
of $\xi$, in fact, result in smaller periods for pulsar B: this produces a 
steep increase of $\dot E_{B}$  which soon would become $\sim \dot E_{A}$ at B's 
light cylinder radius.
Albeit coherence seems indeed to be required in both systems,
no detailed model for coherent radio emission has been put forward so far.
In TT04 some possibilities were suggested but the very model of coherent radio emission
should be revisited possibly in the light of most recent studies on this topic (e.g.
\citealt{fuku04}; \citealt{zl04}).

\cite{hym05} reported that \gcrt \ was not detected in two $\sim 6$~hr images
obtained in 1996--1998 and in $\sim 1$~hr observations in
2002--2003. While no conclusive evidence can be drawn from the shorter observations which
have duration less than the orbital period, the lack of activity in 1996--1998 might
pose a difficulty to the NS+NS binary model, as the same authors state. Here we
only mention that an ``on''--``off'' state may be achieved if star A precesses. For
$\chi_A\sim 0.1$, the wind beam is $\sim 70^\circ$ wide and even a moderate precession
angle can prevent the beam from A from intercepting the 
magnetosphere of B which subtends an angle $\lesssim 10^\circ$. 
Interestingly, the characteristic time $1/\Omega\sim 116  
P^{5/3}M^{-2/3}(1-e^2)\, {\rm s}$ (\citealt{daruf74}) for
changing the system geometry due to geodetic
precession of the spin axis of pulsar A is $\sim 3$~yr, assuming equal
masses ($M=1.3 M_\odot$) for the two stars.

The range of $P_{B}$ and $B_{B}(R_B)$ for which our model holds implies that
neutron star B may be an active radio-pulsar. The required spin down
luminosity of neutron star A (see eq. [\ref{edota}]) is also compatible with it
being a canonical radio-pulsar with $B_{A}\sim 10^{12}-10^{13}$~G and
$P_{A}\sim 0.1-1$ s or a mildly recycled pulsar with $B_A\sim 5\times
10^{9}-5\times 10^{10}$ G and $P_{A}\sim 10-20$~ms. The latter
possibility fits better with the standard scenario for the formation of
a double neutron star binary (e.g. \citealt{vdhdl73}), which
predicts that the system should include a relatively young neutron
star (the one resulting from the second supernova explosion) and a
recycled pulsar (the first born neutron star), spun up to tens of
millisecond period by the short phase of mass transfer from the
evolving companion.

Detectability of the pulsed signal from these two
sources depends on the radio luminosity and favorable orientation of
their beams. Given the position in galactic coordinates ($l=358.9^{\circ}$, 
$b=-0.5^{\circ}$) and for a distance in the range 6--12 kpc, the
available models for the distribution of the ionized components in the
interstellar medium (\citealt{tc93}; \citealt{lc01})
indicate that scattering would prevent detection of any pulsating
radio signal with $P$ shorter than few seconds at the frequency of 330 MHz.
However, observations at frequencies higher than 1 GHz would
preserve the possibility of discovering a $\sim 0.3$~s pulsar,
while observing at frequency $\gtrsim 2$ GHz is required for detecting a 
$\sim 10$~ms pulsar.

An X-ray luminosity coincident with the shock of $L_{X}\approx 10^{32}\
{\rm erg\, s}^{-1}$ should be present, assuming that $\sim 10\%$ of the wind
luminosity intercepted by the magnetosphere is converted into X-rays. However,
the expected flux is extremely low, $\approx 10^{-4}\ \rm{cts\, s}^{-1}$
for {\em XMM}-EPIC and {\em Chandra}-ACIS, assuming a power-law spectrum with index 2
and $N_{H}\sim 10^{22}\, \rm{cm}^{-2}$.
We do not expect substantial emission
in other energy bands, as in the case of \dblpsr, although the younger NS may be associated
with a supernova remnant.

Indirect support
in favor of a double NS system for \gcrt \ could come from further radio
observations of \dblpsr. In fact, evidence that 
the continuous source is unresolved at the arcsec level would rule out
the transparent scenario (TT04), leaving coherent emission from the shock as
the only viable option. If the picture of \gcrt \ proposed here
will be confirmed, one should consider that other similar systems may exist in the
Galaxy. Their search among radio transients with programs analogous to that of
\cite{hym05}, may be of the paramount importance for estimating the coalescence
rate on NS+NS systems, a vital issue in connection with the response of new
generation gravitational wave detectors like VIRGO and LIGO.

\begin{acknowledgements}
We thank an anonymous referee for some helpful comments.
Work partially supported by the Italian Ministry for Education,
University and Research (MIUR) under grant PRIN-2004023189.
\end{acknowledgements}

\newpage

\begin{thebibliography}{}
\bibitem[\protect\citeauthoryear{Burgay et al.}{2003}]{burgay03}
Burgay, M., et al. 2003, \nat, 426, 531
\bibitem[\protect\citeauthoryear{Damour \& Ruffini}{1974}]{daruf74} 
Damour, T., \& Ruffini, R. 1974, Comptes Rendus Acad. Sci. Ser. A, 279, 971 
\bibitem[\protect\citeauthoryear{Cordes \& Lazio}{2001}]{lc01}
Cordes, J.M., \& Lazio, T.J.W. 2001, \apj, 549, 997
\bibitem[\protect\citeauthoryear{Fung \& Kuijpers}{2004}]{fuku04}
Fung, P.K., \& Kuijpers, J. 2004, \aap, 422, 817
\bibitem[\protect\citeauthoryear{Hyman et al.}{2005}]{hym05}
Hyman, S.D., et al. 2005, \nat, 434, 50
\bibitem[\protect\citeauthoryear{Lyne et al.}{2004}]{lyne04}
Lyne, A.G., et al. 2004, Science, 303, 1153
\bibitem[\protect\citeauthoryear{Lyutikov}{2004}]{lyut04}
Lyutikov, M. 2004, \mnras, 353, 1095
\bibitem[\protect\citeauthoryear{Taylor \& Cordes}{1993}]{tc93}
Taylor, J.H., \& Cordes, J.M. 1993, \apj, 411, 674
\bibitem[\protect\citeauthoryear{Turolla \& Treves}{2004, hereafter TT04}]{tt04}
Turolla, R., \& Treves, A. 2004, \aap, 426, L1 (TT04)
\bibitem[\protect\citeauthoryear{van den Heuvel \& de Loore}{1973}]{vdhdl73}
van den Heuvel, E.P.J., \& de Loore, C. 1973, \aap, 25, 387
\bibitem[\protect\citeauthoryear{Zhang \& Loeb}{2004}]{zl04}
Zhang, B., \& Loeb, A. 2004, \apj, 614, L53
\bibitem[\protect\citeauthoryear{Zhu \& Xu}{2005}]{zx05}
Zhu, W.W., \& Xu, R.X. 2005, preprint (astro-ph/0504251)

\end{thebibliography}
\end{document}